\documentclass[%
 reprint,
 amsmath,amssymb,
 aps,
]{revtex4-2}

\usepackage{graphicx}
\usepackage{dcolumn}
\usepackage{bm}
\usepackage{hyperref}
\usepackage[mathlines]{lineno}
\usepackage[position=top, singlelinecheck=false]{subfig}
\usepackage{comment}

\begin{document}

\preprint{APS/123-QED}
\title{A physically-informed sea spray generation model for splashing waves} 

\author{Kaitao Tang}
 \email{kaitao.tang@eng.ox.ac.uk}
\author{Thomas A.A. Adcock}
\author{Wouter Mostert}
\affiliation{Department of Engineering Science, University of Oxford, Oxford OX1 3PJ}

\date{\today}

\begin{abstract}
Large sea spray drops - of up to 2mm in diameter - constitute one of the most uncertain factors controlling the intensification of hurricanes and severe storms because their generation mechanisms are not understood. Wave splashing produces among the largest spray drops, but observational data regarding these drops is difficult to obtain and hence cannot inform current modelling efforts. In this study, we instead propose a sea spray generation function (SSGF) for ocean wave splashing by assembling a model from first principles. First, we introduce the transverse collision of two cylindrical liquid rims as the basic mechanism for drop production. We characterize the resulting drop production in terms of three competing processes: ligament production and merging, drop generation by end-pinching, and gravity which arrests the mechanism. Second, we formulate a theoretical model which explains the drop size distributions produced by the colliding rims and test it against existing experimental and numerical data. Finally, the model can be developed into a full SSGF by incorporating sea state information with relatively few tuning parameters. The model is flexible and can be extended by including related effects such as finite droplet lifetime and secondary breakup. Altogether, our model suggests that wave splashing can efficiently produce numerous secondary droplets, challenging prior assumptions that it is an inefficient generation mechanism for sea spray.
\end{abstract}


\maketitle

The splashing of water is among the most familiar of everyday experiences, appearing in a multitude of activities in science and nature. When two bulks of water collide, such as in breaking ocean waves, they can produce large drops, up to 2mm in size. However, the role of splashing in generating large sea spray drops remains largely unexplored. This is important because large drops may account for the saturation of the sea-surface drag coefficient that occurs in the development of violent storms at sea \cite{Veron2015}. As recent hurricanes show record strengths and growth rates, there is an urgent need for improving the accuracy of air-sea interaction models \cite{Troitskaya2017}, including the parameterisations of sea spray generation rate and size distributions which are still mostly empirical and feature large variability of several orders of magnitude \cite{Veron2015}. This uncertainty reflects the limited knowledge of the sea spray production mechanisms: while spray droplets are known to form through a few major pathways, including surface bubble bursting, wind-driven spume ejection and wave splashing \cite{mueller2014impact, Veron2015, deike2022mass}, detailed experimental and numerical measurements of spray generation are limited. Moreover, no existing SSGF parameterization has taken into account the contribution of splash drops \cite{Veron2015}. Consequently, a general theoretical model for the associated droplet statistics is not yet available \cite{Mostert2021, deike2022mechanistic}. 

In this Letter, we propose from first principles an SSGF for splash droplets from wave breaking. We hypothesise that secondary splashing in breaking waves can be modelled by the transverse collision of two liquid rims. The collision produces an expanding liquid lamella featuring an array of ligaments, which generate individual droplets \cite{tang2024fragmentation}. We show that the resulting droplet size distribution can be captured theoretically depending on the collision parameters with limited reliance on fitting parameters. The model predicts that splash droplets are more prevalent than previously assumed. Moreover, the formulation of the model is flexible and can be extended to include different physical effects or adjusted as new observational data become available in the future.

\begin{figure*}[htbp]
	\centering
		\includegraphics[width=.99\textwidth]{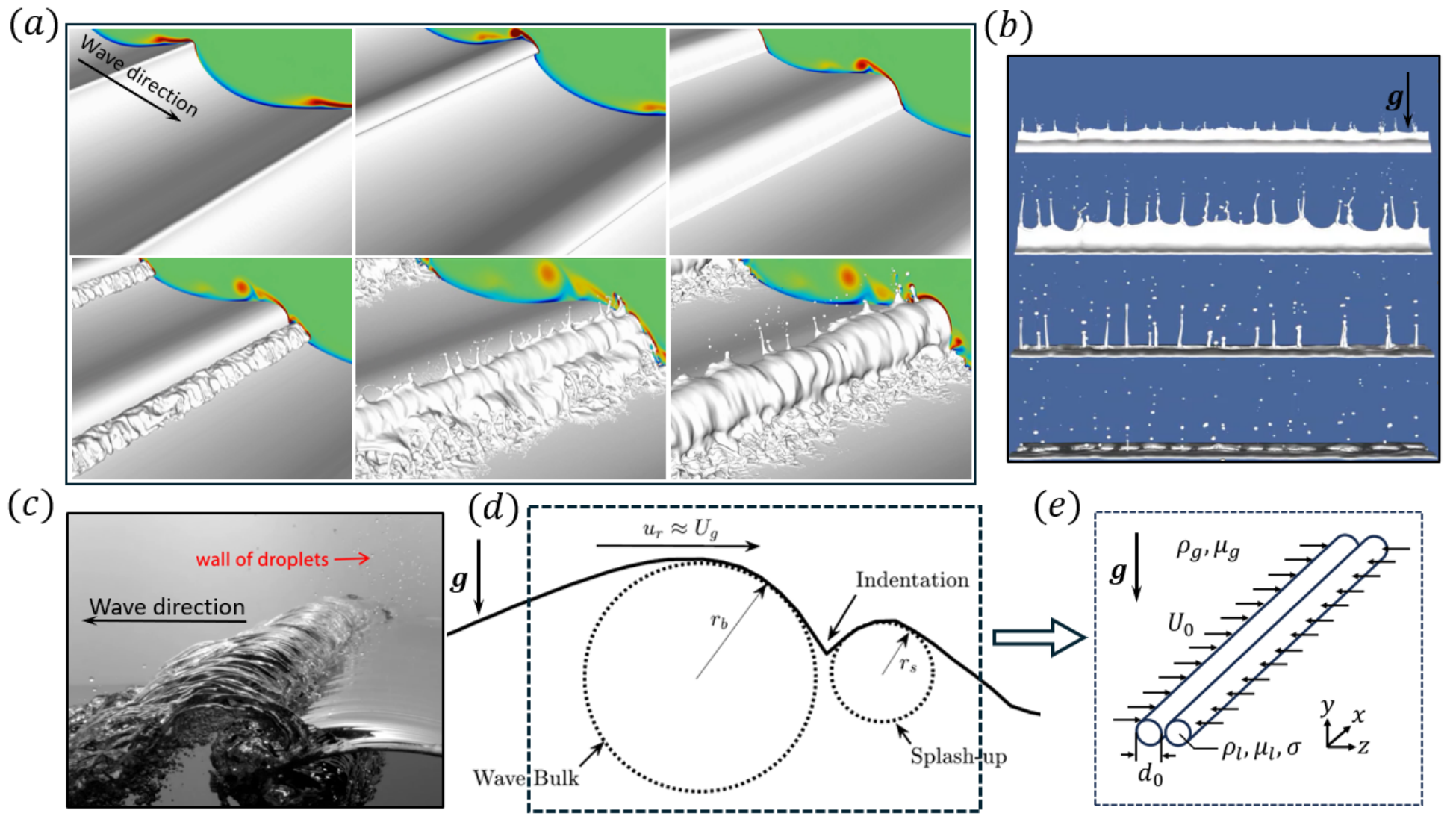}
	\caption{(a): Morphological evolution of a deep-water plunging breaker, with secondary splashing shown in the two frames at the bottom right \cite{Mostert2021}. The wave travels from top left to bottom right in each frame. (b): Side views of a typical rim splashing process. (c): Experiment from Ref.~\cite{erinin2023plunging} showing the generation of secondary splash drops. (d,e): Sketches showing the configurations of secondary wave splashing and rim collision, adapted from Refs.~\cite{erinin2023plunging, tang2024fragmentation}. }
	\label{fig:rim-splash-config}
\end{figure*}

A deep-water surface wave breaking with amplitude $a_b$, wavenumber $k_b$, surface tension $\sigma$ in the presence of gravitational acceleration $g$, as shown in fig.~\ref{fig:rim-splash-config}a, can be characterised by the Bond number $Bo_b \equiv \rho_l g /\sigma k_b^2$ and slope $S_b \equiv a_b k_b$. As the wave steepens and breaks, the overturning jet impacts the water surface and produces a splash-up, which is a bulk of rebounding water partly decelerated by the action of breaking. The remainder of the wave bulk thus collides with the splash-up with a relative speed $u_r=\alpha U_g$, where $U_g = \sqrt{g/k_b}$ is the linear phase speed of the breaking wave. This phenomenon produces a significant number of droplets, which plays a dominant role in the spray statistics \cite{erinin2023plunging}. The phenomenon may be modelled by the collision of two cylindrical liquid rims with diameter $d_0 = \chi a_b$ aligned along the transverse direction, and the relative speed between the two rims is $u_r$, as shown in fig.~\ref{fig:rim-splash-config}d,e and investigated previously in Ref.~\cite{tang2024fragmentation}. Here, $0 \leq \chi, \, \alpha \leq 1$ are dimensionless $O(1)$ coefficients. Gravitational acceleration $g$ acts perpendicular to the horizontal plane containing the two rims, and the Weber and Bond numbers of rim collision are related to the breaking wave parameters as follows:
\begin{equation}
    We \equiv \frac{\rho_l u_r^2 d_0}{\sigma} = \alpha^2 \chi Bo_b S_b, \quad Bo \equiv \frac{\rho_l g d_0^2}{\sigma} = \chi^2 Bo_b S_b^2.
    \label{for:non-dimensional-groups}
\end{equation}
To model the random perturbations on the initial splash-up and wave bulk during wave splashing, we impose a truncated white-noise surface perturbation on the rims \cite{Mostert2021, pal2021statistics, tang2024fragmentation} characterised by $\varepsilon_0$, the non-dimensionalised characteristic perturbation amplitude; and $N_{\rm max}$, the cutoff wavenumber of the perturbation signal spectrum.

Fig.~\ref{fig:rim-splash-config}b provides an overview of the rim splashing phenomenon. At sufficiently large $We$, the two perturbed rims merge and generate a vertically expanding lamella bordered by a thickening rim. The bordering rim is topped by many ligaments ejecting small fragments via the end-pinching mechanism \cite{gordillo2010generation}. In the meantime, these ligaments migrate along the corrugated lamella rim and merge to form thicker ligaments, which causes the gradual increase in the size of emitted splash drops \cite{tang2024fragmentation}. However, gravity causes the lamella base to gradually spread out along the $xz$-plane, followed by the retraction of the lamella sheet. Ultimately, all ligaments are destroyed and secondary droplets fall back to the liquid surface \cite{Veron2015, Mostert2021}.
\begin{figure*}[htbp]
	\centering
	\subfloat[]{
		\label{fig:timescales}
		\includegraphics[width=.92\textwidth]{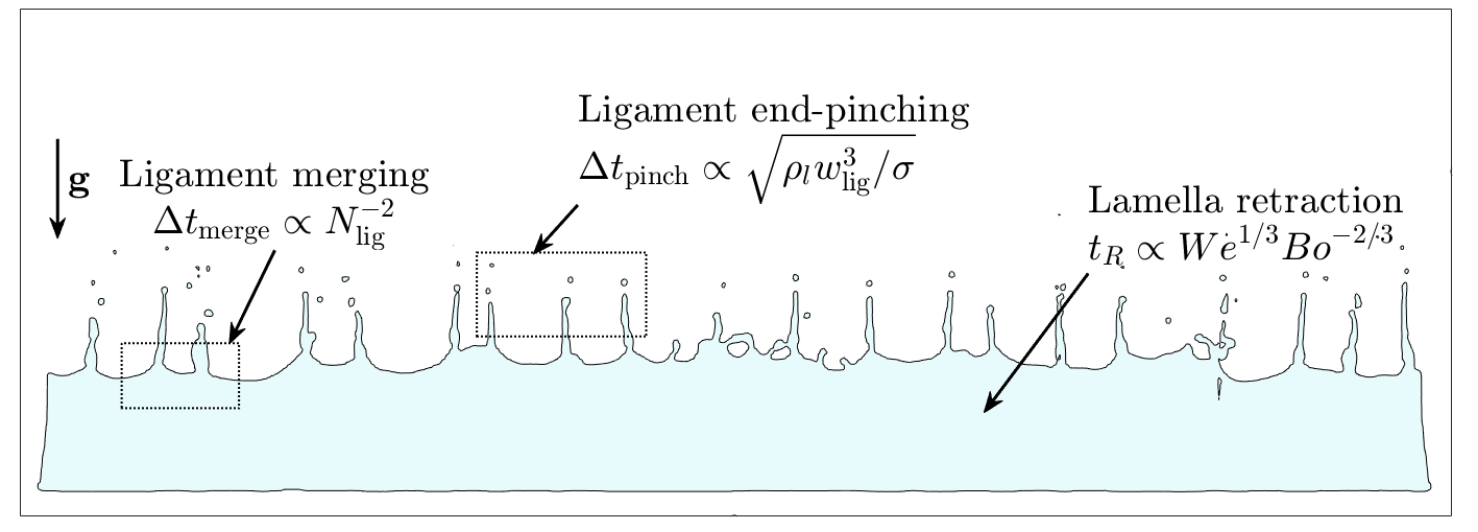}}

    \centering
	\subfloat[]{
		\label{fig:size-pdf-comp-Bo-0}
		\includegraphics[width=.48\textwidth]{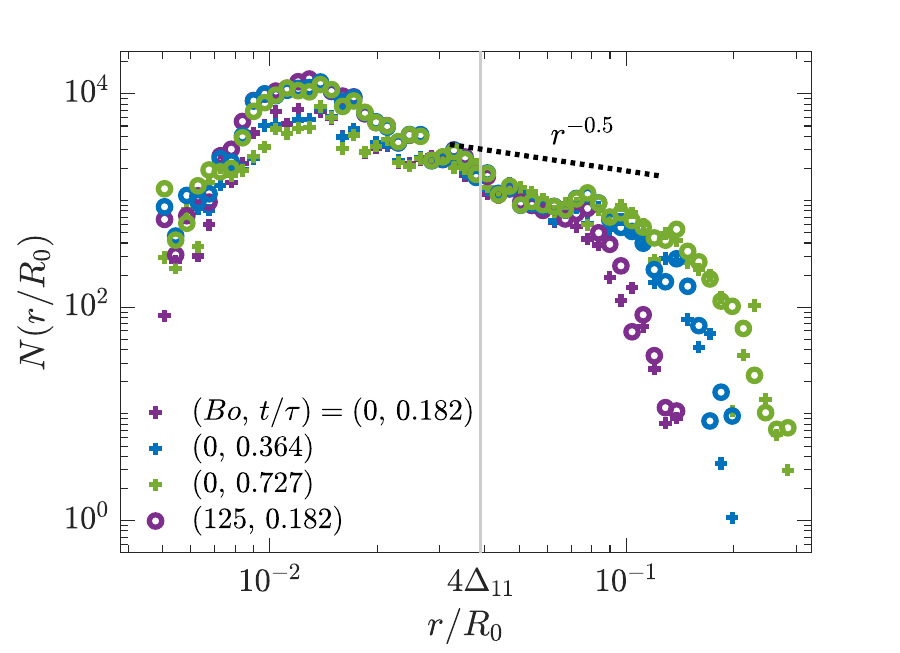}}
    \centering
	\subfloat[]{
        \includegraphics[width=.48\textwidth]{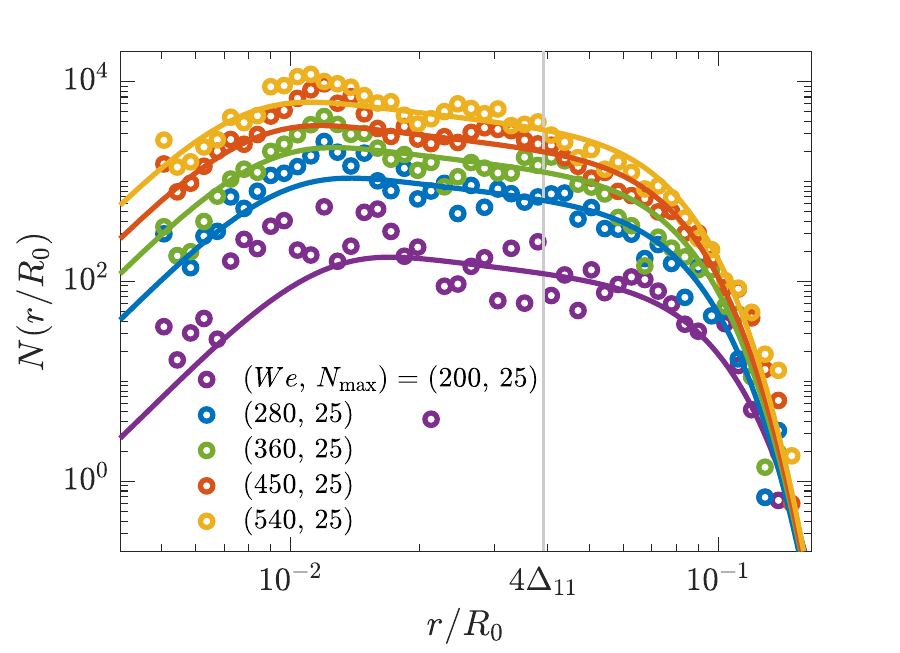}
	   \label{fig:size-pdf-we-nmax-sweep}}
	\caption{(a): Sketch showing all timescales present in the current rim splashing configuration. (b): The evolution of the fragment size distribution of rim splashing over time. Coloured crosses and circles indicate rim splashing with $Bo = 125$ and 0, respectively. The threshold for grid convergence, $r = 4\Delta_{11}$, has been marked in all subsequent plots presenting fragment size distributions. (c): Splash drop size distributions at $t/\tau_{\rm cap} = 0.227$ for different $We$ (scattered points), in comparison with the prediction of \eqref{for:size-dist-model} (solid lines).}
	\label{fig:rim-frag-size-evol}
\end{figure*}

These observations suggest three different timescales present within rim splashing, which we show schematically in fig.~\ref{fig:timescales}: the ligament merging timescale $\Delta t_{\rm merge}$, the ligament end-pinching timescale $\Delta t_{\rm pinch}$, and the lamella retraction timescale $t_R$. The former two timescales have been quantified in our previous work \cite{tang2024fragmentation}, whereas the lamella retraction timescale $t_R$ arises due to gravity and serves as the cutoff timescale of splashing. The rim undergoes a free-fall motion superimposed on the capillary deceleration, the latter of which follows a $t^{1/2}$ power law \cite{roisman2002impact, tang2024fragmentation}. Balancing free fall against lamella expansion yields,
\begin{equation}
    t_R/\tau_{\rm cap} \propto We^{1/3} Bo^{-2/3},
  \label{for:retract-timescale}
\end{equation}
where $\tau_{\rm cap} \equiv \sqrt{\rho_l d_0^3/8\sigma}$ is the capillary timescale.

As splashing terminates at a finite time $t_R$, the unsteadiness of early-time fragmentation plays a significant role, and we expect the fragment size distribution to deviate from quasi-steady droplet production \cite{tang2024fragmentation}. Fig.~\ref{fig:size-pdf-comp-Bo-0} shows the fragment number density $N(r)$ for two different $Bo$ values (0 and 125) and three different times ($t/\tau_{\rm cap} = 0.182$, 0.364 and 0.727), while $We = 280$ and $N_{\rm max} = 60$ for both cases. A vertical line corresponding to droplets of radius $4\Delta_{11}$, below which grid convergence has not been verified, is included. At fixed times, finite $Bo$ values do not modify the fragment size distribution for $r \geq 4\Delta_{11}$, since gravity does not affect the ligament merging and end-pinching dynamics.

While the quasi-steady $r^{-1/2}$ power law derived in Ref.~\cite{tang2024fragmentation} is observed here for $4\Delta_{11} \leq r/R_0 \leq 10^{-1}$, an unsteady model is required to fully describe the evolving right tail of the droplet size distributions $N(r/R_0)$, which is linked to the distribution of the width $w$ of rim ligaments via the ongoing end-pinching process \cite{wang2018unsteady}. We assume that the distribution of ligament widths $w$ abides by a time-dependent Gamma distribution of order $m$ \cite{Villermaux2007},
\begin{equation}
    p \left[\frac{w}{\bar{w}(t)} \right] = \frac{m^m}{\Gamma(m)} {\left [ \frac{w}{\bar{w}(t)} \right]}^{m-1} e^{- \frac{mw}{\bar{w}(t)}},
    \label{for:width-dist-gamma}
\end{equation}
where $\bar{w}(t) \propto \sqrt{t/\tau_{\rm cap}}$ is the average ligament width \cite{tang2024fragmentation}. Other statistical models aside from \eqref{for:width-dist-gamma} may be used; for example, the log-normal distribution \cite{tang2022bag} produces similar results to the below (see Supplementary Information~D).

Now consider all pinch-off events for ligaments with width $w(t)$ having occurred for $t_0 \leq t \leq T$. The onset time for ligament pinch-off $t_0$ can be estimated as the duration of rim indentation closure after impact,
\begin{equation}
    \frac{t_0}{\tau_{\rm cap}} \approx \frac{\varepsilon_0 d_0}{U_0 \sqrt{\frac{\rho_ld_0^3}{\sigma}}} = \frac{\varepsilon_0}{\sqrt{We}}.
\end{equation}
As ligaments pinch off at a rate of $1/\Delta t_{\rm neck} \propto w^{-3/2}$ \cite{tang2024fragmentation}, the total number density of fragments with radius $r$ at a given time $t$ is therefore
\begin{widetext}
\begin{gather}
    N(r, \, t) \propto \frac{1}{\Delta t_{\rm neck}} \int_{t_0}^t N_{\rm lig} (t') \cdot p \left[\frac{w}{\bar{w}(t')} \right] {\rm d}t' 
    = C \frac{m^3}{\Gamma(m)} {\left( \frac{r}{R_0} \right)}^{-\kappa} \left( \Gamma_{\rm inc} \left[m-2, \, m\frac{r}{\bar{r}(t)} \right] - \Gamma_{\rm inc} \left[m-2, \, m\frac{r}{\bar{r}(t_0)} \right] \right),
    \label{for:size-dist-model}
\end{gather}
\end{widetext}
where $\Gamma_{\rm inc}$ is the incomplete Gamma function. Here we utilise the observation $\bar{r}(t) \approx 0.7\bar{w}(t)$ established in Refs.~\cite{tang2024fragmentation, wang2018unsteady} for ligament end-pinching. For small fragment sizes $r/R_0 \ll 1$ and pinch-off onset time $t_0 / \tau_{\rm cap} \ll 1$, \eqref{for:size-dist-model} reduces to a power-law model $r^{-\kappa}$, whereas the right tail at large droplet sizes is now shown to arise from the non-uniform distribution of ligament widths. While $\kappa=1/2$ in the theoretical model above, we show in Supplementary Information B that it may take different values in the range $1/2 \leq \kappa \leq 2$, to allow for variations in the splashing process. Such variations may arise from details not included in the basic physical model, such as collision between asymmetric rims, effects from short-crestedness, and so on. Whatever the precise value of $\kappa$, the functional form of \eqref{for:size-dist-model} resembles final fragment size distributions for other types of spray, see for example Refs.~\cite{Berny2021, deike2022mechanistic}. 

In fig.~\ref{fig:size-pdf-we-nmax-sweep} we compare \eqref{for:size-dist-model} with the fragment size distributions for simulations with $200 \leq We \leq 540$ and $N_{\rm max}=25$. Increasing $We$ leads to an overall increase in the fragment number density $N(r/R_0)$. Here, \eqref{for:size-dist-model} matches excellently with the numerical size distributions at different $We$ values, where the integral prefactor in \eqref{for:size-dist-model} is fitted as $C = 1.29 {We}^{2.6}$ and $m$ is fixed at 12. The power-law decay region where $N(r) \propto r^{-1/2}$ is observed extends to $r/R_0 = 0.01$, below which $N(r)$ starts to decrease. Our size distribution \eqref{for:size-dist-model} successfully captures this decrease at small fragment sizes, suggesting its possible origin in the initial growth of transverse ligaments before the first pinch-off events. We have also validated \eqref{for:size-dist-model} against available wave breaking data (see Supplementary Information B).

The development of the fragment size distribution \eqref{for:size-dist-model} for individual splashing events allows prediction of splash drop production within realistic sea states, where breakers of different length and time scales are present simultaneously. The sea spray generation function, denoted $dF/dr$, measures the spray production rate per unit ocean surface area and drop radius increment, and aids the calculation of air-sea mass, momentum and energy exchange \cite{mueller2014bimpact, Veron2015}. However, currently available SSGFs feature a considerable range of scatter for large spray drops with radii larger than $10 {\rm \mu m}$, and none explicitly addresses the production of splash drops \cite{mueller2009sea, Veron2015}. Here we estimate an SSGF of splash drops for the first time.

\begin{figure*}[htbp]
    \centering
    \subfloat[]{
    \label{fig:ssgf-devel}
    \includegraphics[width=.48\textwidth]{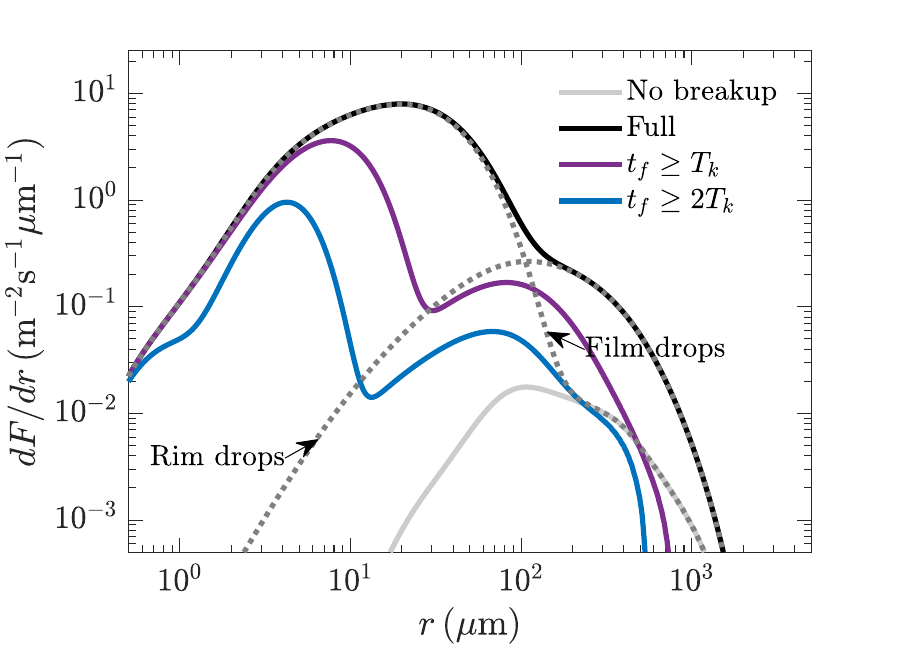}}
    \centering
    \subfloat[]{
    \label{fig:ssgf-comp}
    \includegraphics[width=.48\textwidth]{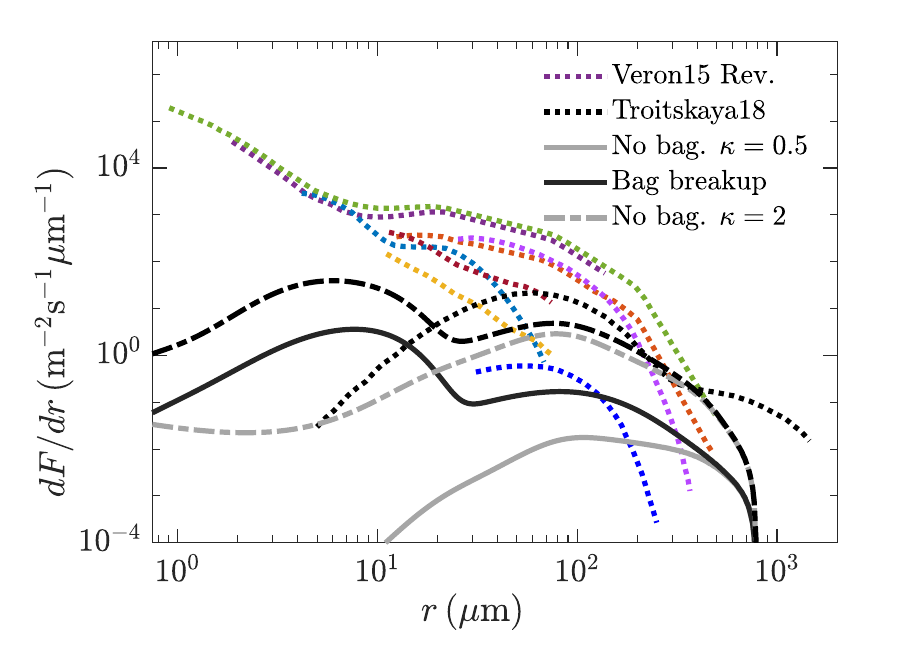}}
    \caption{
    (a): Predicted SSGFs without (grey solid line) and with (black solid line) secondary bag breakup for splash drops, with contributions from bag film and rim drops shown for the latter. (b): Comparison of our SSGFs with those reviewed by Veron \cite{Veron2015} (coloured dotted lines) and recently proposed by Troitskaya \emph{et al.} \cite{Troitskaya2018} (black dotted lines). Sources for previous SSGF models reviewed by Veron \cite{Veron2015} are listed below. Green: Ref.~\cite{fairall1994effect}; dark purple: Ref.~\cite{andreas1992sea}; light purple: Ref.~\cite{fairall2009investigation}; orange: Ref.~\cite{andreas1998new}; yellow and dark brown: Ref.~\cite{pattison1999production}; dark blue: Ref.~\cite{smith1993marine}; light blue: Ref.~\cite{mueller2009sea}.}
    \label{fig:ssgfs}
\end{figure*}

Since splashing is intrinsically linked to the appearance of a breaking wave, the omnidirectional breaking wave distribution $\Lambda (k)$ \cite{phillips1985spectral, romero2019distribution, deike2022mass} is a promising agent to capture the SSGF. This function describes the total length of breaking crests per unit sea area, as a function of the wavenumber $k$. Assuming that breaking crests at all wavelengths are sufficiently strong to undergo secondary splashing, the SSGF for splash drops can be computed as follows,
\begin{equation}
    \frac{dF}{dr} (r) = \int_{k_{\rm min}}^{k_{\rm max}} \frac{\Lambda(k)}{l_0 t_b(k)} N(r, \, t_R) dk.
    \label{for:ssgf-integral}
\end{equation}
Here we estimate $\Lambda(k)$ based on the saturated breaking wave field investigated by Wu \emph{et al.} \cite{wu2023breaking}. 
The scaling factor $l_0$ accounts for the spanwise splashing length over which the fragment size distribution $N(r)$ is computed. In our simulations $l_0$ is set as ten times the rim diameter $d_0$. According to the analogy between rim and wave splashing introduced in Eq.~\eqref{for:non-dimensional-groups}, $d_0 = \chi a_b = \chi s_b/k$. $t_b(k) \equiv 1.12\pi/\sqrt{gk}$ is the breaking time for the wave component with wavenumber $k$ \cite{romero2019distribution, wu2023breaking}. More detailed information about the calculation of $dF/dr$ is available in Supplementary Information~C.

The SSGF calculated directly from \eqref{for:ssgf-integral} is shown with the grey solid line in fig.~\ref{fig:ssgf-devel}. The shape of this SSGF closely resembles that of the fragment size distribution, featuring a plateau where $dF/dr \propto r^{-1/2}$ flanked by two fall-off tails. Notably, the plateau spans over a radius range of $100 \, {\rm \mu m} \leq r \leq 300 \, {\rm \mu m}$, overlapping with the size range for typical ocean spume drops and supporting the speculation of Veron \cite{Veron2015} regarding whether splash drops may be roughly included in the spume population.

While the SSGF based on \eqref{for:size-dist-model} features a small total number of fragments \cite{Veron2015}, high-amplitude breaking waves may generate sufficiently large splash drops, which fragment to produce large numbers of smaller secondary droplets and modify the shape of the fragment size distribution $N(r)$ despite their rare occurrence. This may be realised via droplet aerobreakup, where the fragmentation of thin bag films is known to introduce many fragments significantly smaller than the original droplet \cite{Troitskaya2018, tang2022bag}. Fig.~\ref{fig:ssgf-devel} shows the modified SSGFs incorporating secondary aerobreakup of splash drops in solid black line. This version of the SSGF features two peaks characteristic of film and rim drops produced from bag breakup events \cite{Troitskaya2018}. Compared with the previous version of SSGF without secondary aerobreakup, the right tail does not show any obvious shift, while the number of small fragments increases significantly due to bag film fragmentation. Other modifications to the SSGF can be made: for example, recognising that large droplets have a short lifetime allows for a filter to be included of the form $t_f \geq \nu T_k$ (where $\nu$ is a constant), which shifts the SSGFs further towards smaller fragment sizes as short-lived large drops are filtered out, while rendering the characteristic two peaks of bag breakup more distinctive. The SSGF models incorporating aerobreakup of splash drops compare reasonably with available empirically fitted models \cite{Veron2015} in fig.~\ref{fig:ssgf-comp}. We note that this match with previous results can be further improved if we set $\kappa = 2$ rather than 1/2, which would align the predicted fragment size distribution closer to the upper envelope of the aforementioned models. Overall, our results suggest that once the secondary fragmentation of large splash drops is included, wave splashing may be an indirect but nevertheless efficient spray generation mechanism (though note Ref.~\cite{mueller2014impact}).

In conclusion, we have numerically investigated rim splashing under gravity as a model for the wave splashing phenomenon. Gravity introduces a global cut-off timescale for the ongoing ligament merging and end-pinching competition, which determines the final size distribution of splash drops. To account for the early-time broadening of the fragment size distributions, we propose a theoretical model which compares favourably with available data. This further enables the development of an SSGF accounting for splash drop production for the first time. The prediction of this model suggests that wave splashing can potentially generate more spray drops than previously expected as large splash drops undergo aerobreakup and produce numerous small film drops. Our results complement the ongoing development of SSGFs centred around contributions from film and jet drops arising from surface bubble bursting \cite{Berny2021, deike2022mechanistic}, and thus pave the way for the development of more accurate ocean spray parameterisations to be incorporated in future global climate models \cite{mueller2014bimpact}. 

Note finally that the SSGF models presented in this work are best construed as upper bounds for the production rate of splash drops, as many open questions remain regarding wave splashing. These include the characteristic wavelength and amplitude of the initial splash-up surface perturbations, and the splashing mechanism in open-ocean conditions as the assumption of two-dimensionality breaks down \cite{mcallister2024three}. Further experimental and observational studies are therefore required. But the SSGF proposed above presents a robust, physics-based core model that can be appropriately modified and further improved to capture the relevant phenomena as they become better understood and observed in the ocean environment, for example, turbulence in the ocean-atmospheric boundary layer and microphysics of individual spray drops \cite{mueller2014impact}.

The authors thank EPSRC for accessing the UK supercomputing facility ARCHER2 via the UK Turbulence Consortium (EP/R029326/1), and the University of Oxford for using the Advanced Research Computing (ARC) facility. K.T. is supported by the University of Oxford (Departmental Studentship No. 1411916) and the Natural Environment Research Council (UKRI Grant No. 1271: Contaminated Ocean Spume).

\bibliography{main}

\end{document}